# CHRONEX-US: City-level historical road network expansion dataset for the conterminous United States

[1,2]Johannes H. Uhl, [3]Keith A. Burghardt, [4]Stefan Leyk

[1]Institute of Behavioral Science, University of Colorado Boulder, Boulder (CO), USA.
[2]Joint Research Centre, European Commission, Ispra (VA), Italy.
[3]Information Sciences Institute, University of Southern California, Marina del rey (CA), USA.
[4]Department of Geography, University of Colorado Boulder, Boulder (CO), USA.
Corresponding Author: johannes.uhl@colorado.edu

**Abstract:** Geospatial datasets on the long-term evolution of road networks are scarce, hampering our quantitative understanding of how the contemporary road network has evolved over the course of the 20[th] century. However, such information is crucial to better understand the dynamics of road network growth and expansion, and to shed light on the consequences of (sub-) urbanization processes, such as increasing mobility, traffic congestion, land take and transportation inequality. Herein, we describe CHRONEX-US ("City-level historical road network expansion dataset for the conterminous United States"), a geospatial vector dataset reporting estimates of the construction year for each road segment in densely and semi-densely built-up spaces within 693 core-based statistical areas (i.e., Metropolitan and Micropolitan statistical areas) in the conterminous US. CHRONEX-US is based on the USGS National Transportation Dataset (NTD), integrated with the historical settlement compilation for the US (HISDAC-US). CHRONEX-US reports model-based construction epoch estimates for local residential road network segments within urban and peri-urban areas, using different model-based scenarios. The vector data inherit topological integrity from the NTD data allowing for routing and other connectivity-based analyses within temporal strata of urban road networks. CHRONEX-US vector geometries are attributed with the US Census Bureau's MAF/TIGER Feature Class Code (MTFCC), enabling stratification of the data by road category. CHRONEX-US is available at https://doi.org/10.6084/m9.figshare.28644674.

## 1. Introduction

Geospatial data on the historical extent and the structure of road and other transportation networks and on their evolution over time are critical to understand urban growth, the relationship between road network structure and economic features (Jaworski et al., 2020), car prevalence (Barrington-Leigh and Millard-Ball, 2020), and other features of evolving cities. Moreover, these data could help us predict the future evolution of roads. These data, however, are difficult to acquire, especially at a national level, becaused data collection requires (a) a suitable data source, such as historical maps or atlases, and (b) either expert knowledge to use state-of-the-art computer vision models to extract such information (e.g., Saeedimoghaddam and Stepinski 2020, Chiang et al. 2020, Avci et al 2022, Jiao et al. 2022, Uhl et al. 2022, Jiao et al. 2024, Sertel et al. 2024), or (c) significant amounts of manual work for data digitization (e.g., Thévenin and Mimeur 2023, Turner et al. 2023). Alternative approaches to produce historical models of (urban) road network data include the use of ancillary data describing processes that are presumably coherent to the evolution of road networks. For example, building construction year data have been used, assuming that roads and buildings are approximately built in the same epoch (Barrington-Leigh and Millard-Ball 2015), historical settlement footprints (Barrington-Leigh and Millard-Ball 2020, Burghardt et al. 2022) or multi-temporal urban/rural classifications allowing for the extraction of historical urban road networks (Boeing 2021). Herein, we describe a dataset produced by such a model, relying on historical settlement extents, annotating each present-date road network segment with several construction epoch estimates derived from density-based urban and peri-urban settlement footprints. This dataset has served for several scientific studies in the past. By making these data publicly available we provide a database for analyzing road network evolution over the long term (1900 to 2020), for dense urban and peri-urban areas in almost 700 U.S. metropolitan and micropolitan statistical areas.

## 2. Data sources

CHRONEX-US (Uhl et al. 2025) is based on geospatial, linear vector data from the United States Geological Survey (USGS) National Transportation Dataset (NTD, USGS 2023). NTD data are based on road geometries from US Census TIGER/Line data, enriched with data from HERE road data (USGS 2023). Data was downloaded in 2019, representing the road network approximately in 2018. In order to estimate road network age, we used gridded settlement data from the Historical Settlement Data Compilation for the US (HISDAC-US, Leyk & Uhl 2018, Uhl et al. 2021), and a derived product from HISDAC-US, which is the historical, generalized built-up area dataset (GBUA, Uhl & Burghardt 2022).

HISDAC-US is a compilation of gridded aggregates derived from property records from the Zillow Transaction and Assessment Dataset (ZTRAX), a US-wide harmonized dataset of property data, containing the location, indoor area, and construction year of buildings corresponding to over 150 million properties in the US (Zillow 2025). HISDAC-US provides several gridded, multitemporal datasets, such as the HISDAC-US Built-up Property Locations (BUPL) dataset (Uhl and Leyk 2020), providing a proxy measure of historical building densities at 250m resolution in 5-year intervals from 1810 to 2015, and gridded estimates of historical building footprint area (BUFA; Uhl and Leyk 2022a). HISDAC-US BUFA is an integrated product of property locations from ZTRAX, attributed with building construction year information, and building footprint geometries from Microsoft's US building footprint dataset (Microsoft 2018), aggregated into 250m grid cells, in 10-year intervals from 1900 to 2015. Moreover, we used vector polygon data delineating core-based statistical areas (CBSAs) to partition the data by CBSA (US Census Bureau 2019).

## 3. Methodology

### 3.1. Delineation of historical urban and peri-urban spaces in US core-based statistical areas

CHRONEX-US is based on generalized, historical extents of built-up areas from the historical, generalized built-up area dataset (GBUA). The GBUA dataset has been created by generating focal built-up surface density layers from gridded estimates of historical building footprint area (BUFA) from the HISDAC-US BUFA dataset (250m resolution, at 10-year intervals from 1900 to 2015) using a circular focal window of 1 km radius. For each year, we then followed previous work (Leyk et al., 2018) to delineate the main settlement footprints by excluding areas of <5% built-up surface density, excluding scattered rural settlements. The remaining higher-density grid cells were then segmented to obtain vector objects for each contiguous group of cells. For each vector object, we calculated the total number of buildings within, using the HISDAC-US BUPL dataset of the respective year. For each year, we ranked the segments by the number of buildings within each CBSA. Finally, we retained the upper 10% of these ranked segments, representing high-density patches, thus  excluding extremely small settlements. This process allowed to delineate dense urban spaces in each CBSA, consistent over time. These depictions of historical urban extents have been used to assess road network structure over time (Burghardt et al. 2022),identify urban scaling laws (Burghardt et al. 2024a) as well as explore long-term growth patterns of urban infrastructure (Burghardt et al. 2024b).

## 3.2. Derivation of baseline modelled road age estimates (Model 1)

The model to estimate road network age is based on three assumptions:

**1. Coherence of the evolution of buildings and roads.** In order to estimate the construction epoch of roads, we assume coherence of the evolution of buildings and of the roads "connecting" buildings: If the majority of buildings in a given neighborhood were built in the epoch T, it can be assumed that the local, residential road network was built approximately in the same epoch T as well. This reflects common land development practice, in particular in countries such as the United States, where most of the modern land development occurred in the last two centuries.

**2. The dominant age of the contemporary building stock reflects the period of first development.** We assume that the "original" building stock is still partially existing, and the age of the building stock in a given neighborhood allows for inferring on the time period in which the neighborhood has been developed (Uhl and Leyk 2022b). This is also a fundamental assumption underlying HISDAC-US. This assumption neglects processes of building stock renewal or shrinkage, which, however, are assumed to introduce minor bias only.

**3. Road network stability and "growth-only" assumption.** We assume that the only type of change that a local, residential road network experiences, is expansion: A road network, once it is implanted, is not removed, and its geometric layout is persistent over time. This might break down, for example, when roads are being depaved, or when 4-way intersections are replaced by roundabouts. Broadly, however, we expect this assumption to hold, consistent with prior work (Barrington-Leigh and Millard-Ball 2015, Boeing 2021, Meijer et al. 2018).

The GBUA dataset reflects major development in a spatially aggregated way, in time steps of 10 years. Thus, we overlaid the contemporary NTD road network vector geometries of 2018 to the historical settlement polygons from the GBUA dataset, and attributed each road network segment with time period in which the land was likely to be developed: For example, if a road network segment overlaps the GBUA extent from 1920, it received a lower age estimate of 1910, and an upper age estimate of 1920, as the land was likely developed between 1910 and 1920. If a road network segment overlaps the GBUA extent from 1900, the land was developed any time before 1900; in this case, we set the lower estimate to 0 and the upper estimate to 1900. This is the baseline model (Model 1). It is based on the GBUA extents, which are based on built-up surface densities computed within focal, circular windows of 1km radius. Hence, the GBUA extents represent a generalized estimate of the historical urban fringe, but shifted "outwards" by 1 km. The advantage of this method is that the temporal strata of road networks derived from these age estimates are highly connected and do not contain disconnected parts, favoring routing and connectivity-based analyses. However, this procedure may cause that road network segments in that 1-km band along the fringe are attributed with a built estimate that is too early while the road may have been built later. While this effect is of systematic nature, and likely does not affect analytical outcomes significantly, we provide alternative models that partially mitigate this problem.

## 3.3. Derivation of alternative models (M2, M3)

While model M1 represents generalized and topologically intact historical depictions of the road network, it may overestimate the road network age (i.e., estimate the road construction epoch as too early). Thus, we buffered the GBUA extents "inwards" by applying a buffer distance of -500m (model M2) and -1000m (model M3), causing a "shrinkage" of the historical settlement extents. This shrinkage mitigates the effect of road network age overestimation, as it removes the 1-km buffer produced by the focal density operations applied to the gridded settlement data. However, it produces more segregated depictions of the historical road network, impeding connectivity-based analyses as it creates more disconnected road network segments and parts. Figure 1 shows example depictions of the historical road network based on the three models and the standard deviation between them, calculated per road segment. Meanwhile, Figure 2 illustrates the different road construction epoch estimates combined for the three models. These combined model outcomes (e.g., earliest, latest, or average construction epoch) estimated from the three models allow for variations while better preserving topological integrity (Figure 2).

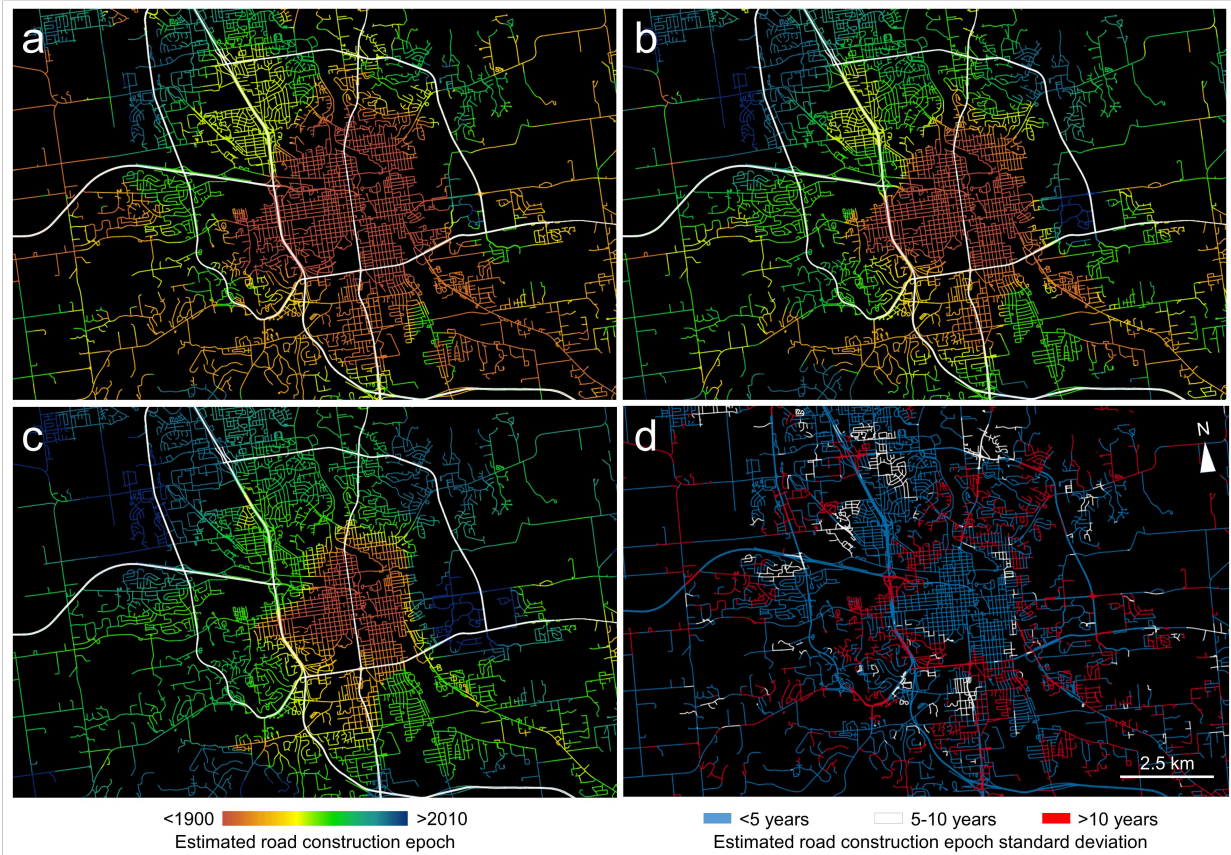

**Figure 1. CHRONEX-US data shown for the city of Rochester, Minnesota: Road construction epochs (lower bounds) estimated from (a) Model 1, (b) Model 2, and (c) Model 3; Panel (d) shows the standard deviation of the three models calculated per road network segment. Roads colored in white in (a)-(c) show highways and other major roads for which the assumptions underlying CHRONEX-US are not valid.**

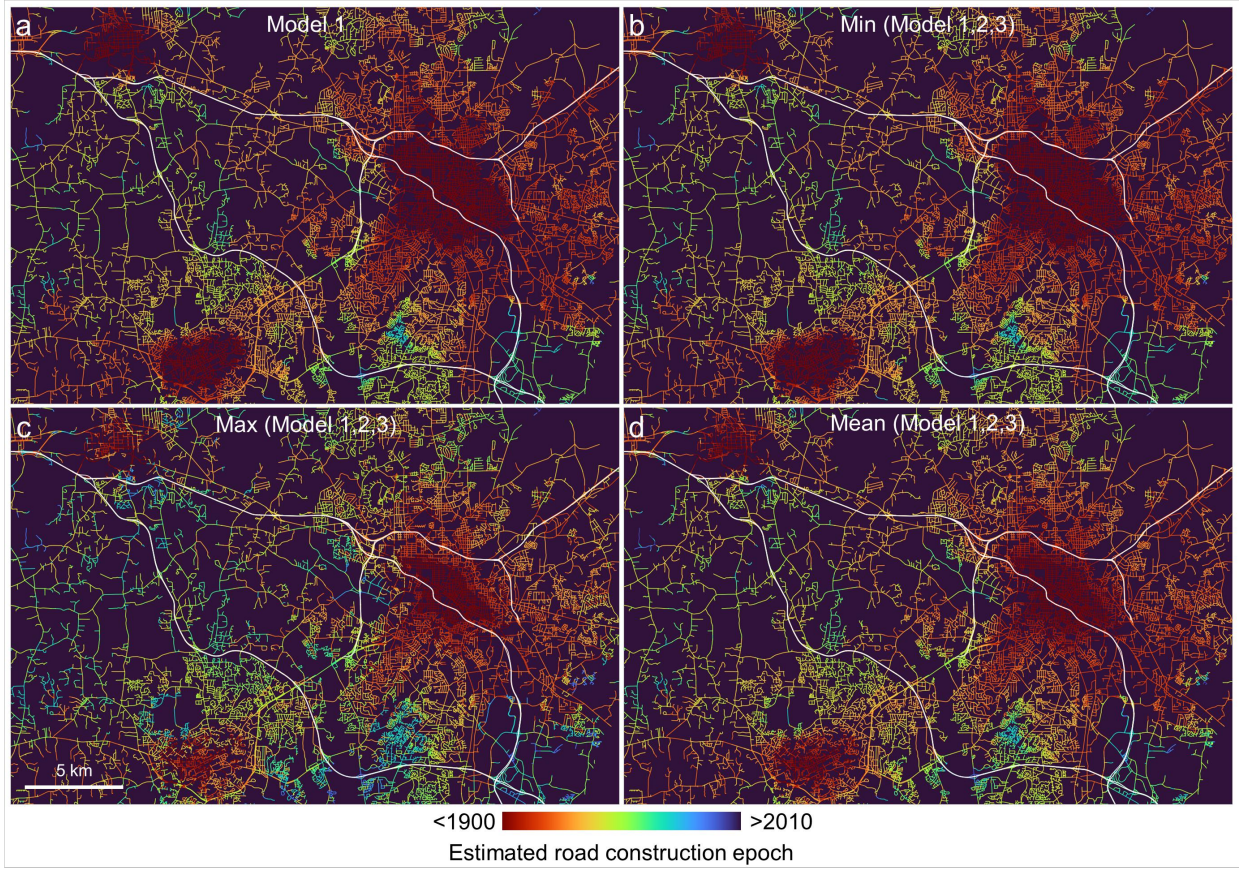

**Figure 2. Comparison of the cross-model road construction epoch estimates. (a) Model 1 results (lower bounds), (b) minimum (i.e., earliest), (c) maximum (i.e., latest), and (d) mean construction epoch estimated per segment by the three models. Data shown for the Durham, Hillsborough, and Chapel Hill region (North Carolina). Primary roads, where model assumptions do not apply, are shown in white.**

## 3.4.    Evaluation

A quantitative assessment of the quality of the historical road network models is challenging due to the lack of historical reference data. To qualitatively assess the plausibility of historical road network depictions in CHRONEX-US, we visually compared early epochs of the data with scanned and georeferenced historical maps from the USGS Historical Topographic Map Collection (HTMC; USGS 2025). Some of these comparisons are shown in Figure 3. As can be seen, CHRONEX-US historical road network depictions approximate the road network shown in the maps relatively well. While the road network footprint may differ, the historical depictions slightly overestimate but capture most parts of the road network shown in the historical maps. One exception is the Albany (New York) example, where CHRONEX-US considerably overestimates the road network of 1900. This is likely an effect of the urban area delineation method, as described above, which may yield inaccurate results in areas that were already densely settled in the early 1900s in both urban and peri-urban or rural areas. In these areas, the thresholds applied to produce the GBUA delineations may not be optimal. Thus, we advise the user to be cautious when using CHRONEX-US in the Northeast of the U.S., in particular in the early-developed New England states. However, this issue does not seem to affect overall trends derived from the data, as the comparison to analytical results (i.e., temporal trends) based on other modeled historical road network data has shown (Burghardt et al. 2022).

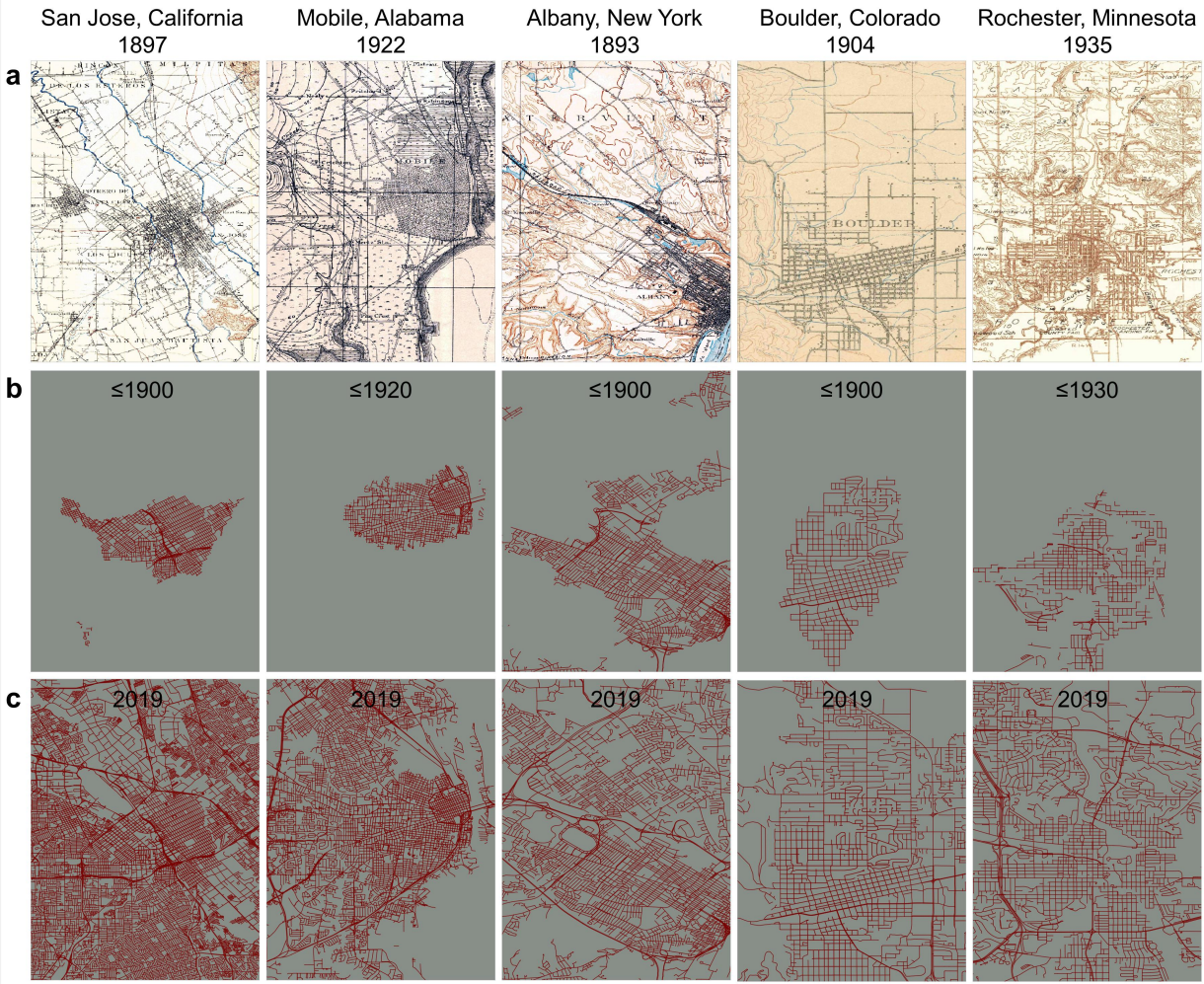

**Figure 3. Historical road network models from CHRONEX-US visually compared to historical topographic maps from 1893 to 1935. (a) USGS HTMC scanned and georeferenced topographic maps (b) CHRONEX-US model-3 estimates for the temporally closest epoch, and (c) the 2019 road network for comparison. Data source (a): USGS 2025.**

Moreover, we quantify the uncertainties introduced and driven by the model selected. In CHRONEX-US, we report a range of age estimates for each road network segment in the GBUA domain as estimated by each of the models. This includes the lower and upper bound as well as the minimum, maximum, and mean of the estimated construction epoch at the road segment level. To quantify the variation across the modeled estimates and the effects on overall road network statistics (i.e., total km road over time), we compare the time series of road length estimated by the different models, in absolute terms (Figure 4a,b) and in relative terms (Figure 4c). We observe that the model choice affects the total estimated road length (Figure 4a), which depends on the spatial extents of the different model domains: The M3 model (inwards-buffered GBUA extents by 1 km) represent the most conservative estimates of the historical road network, as the delineations represent the most constrained delineation of the urban fringe. The combined construction epoch estimates across all models show less variation, and indicate a clear trend of linear urban road network growth up to the 1980s, followed by more moderate growth, with a decreasing growth rate towards 2020 (Figure 4b). When normalizing the time series shown in Figure 4a,b, we observe strong agreement regarding the relative growth of the modeled road networks in recent decades (i.e., narrow inter-quartile ranges), whereas the dispersion between normalized model estimates is higher in earlier epochs, indicating increased levels of uncertainty as we go back in time (Figure 4c). The low estimates of the total urban road network length in the M2 and M3 models (Figure 4a) imply that for analyzing more recent epochs, users should not rely on M2 or M3, but rather use M1 or the cross-model estimates (Figure 4b).

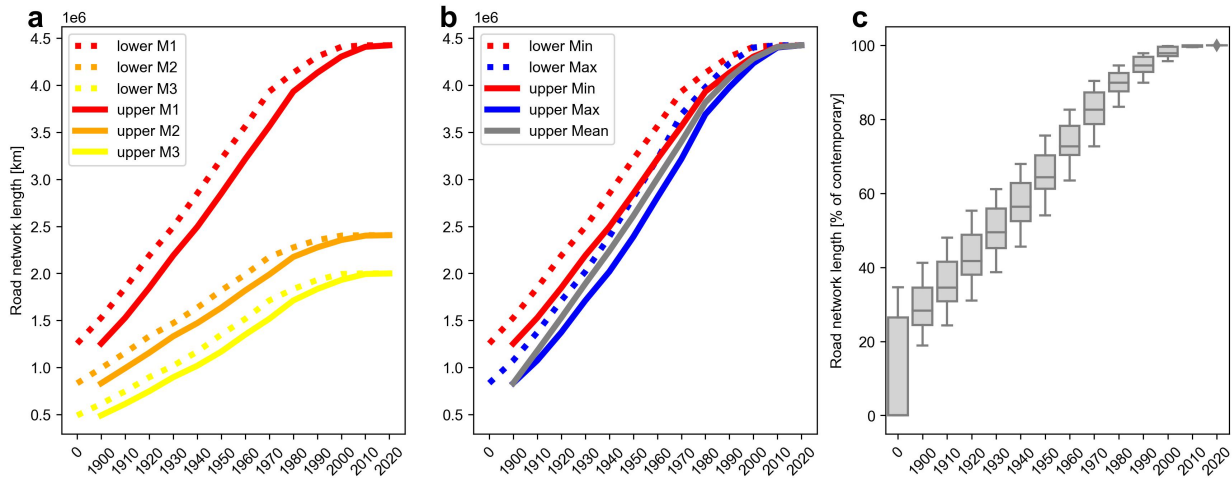

**Figure 4. Total km road over time as modelled by the different model variants in CHRONEX-US. (a) km road over time based on the lower and upper bounds of the construction epoch estimated by the three models (b) trends derived from the road-segment level min, max, mean statistics across the three model estimates, and (c) boxplot of normalized estimates across the 11 models shown in (a) and (b), measured in % of the 2020 total road network length. Note that the mean of the lower estimates across the three models is not reported, as the lower bounds are denoted as 0 and thus, the mean value is meaningless.**

# 4. Data records

CHRONEX-US is available at https://doi.org/10.6084/m9.figshare.28644674 as a ZIP file containing 693 GeoPackage (.GPKG) geospatial vector polyline data files, named "chronex_us_<CBSAID>.gpkg", with <CBSAID> being the 5-digit identifier (GEOID) of the CBSA as defined by the US Census Bureau (see US Census Bureau 2019).

Each GPKG contains the NTD road network vector data (polyline) clipped to the GBUA 2015 extents. Data are provided in projected coordinates, i.e., in the local UTM projection predominant in each CBSA, facilitating distance- and orientation-based road network analyses. The attributes are:

- MTFCC_CODE: The US Census Bureau's MAF/TIGER Feature Class Code (MTFCC; see https://www2.census.gov/geo/pdfs/reference/mtfccs2019.pdf).
- yr_lower_M<1,2,3>: Earliest year of the estimated construction epoch according to model 1,2,3
- yr_upper_M<1,2,3>: Latest year of the estimated construction epoch according to model 1,2,3
- yr_<lower,upper>_<min,max>: Min. and max. of the earliest and latest year of all models.
- yr_upper_<mean,std,range>: Further statistics of the three model outputs. The standard deviation (std) and the range can be used to measure agreement of the three models.

NoData values of the construction epoch estimates indicate that the road network segment was not included in the specific model (i.e., not overlapping with any GBUA polygon after buffering). A lower year of "0" indicates an estimated road construction year <1900. Due to this encoding, the mean, standard deviation and range are only reported for the upper bound and not for the lower bound of the estimated construction epochs.

# 5. Limitations

CHRONEX-US has a number of limitations. We list a few below:

- CHRONEX-US provides model-based estimates of local/residential road network age per road segment, based on historical, spatially generalized delineations of dense- and semi-dense built-up areas, mimicking historical urban and peri-urban settlement extents. CHRONEX-US inherits data incompleteness from HISDAC-US, resulting from property records or missing building construction year attributes in ZTRAX. Several states including North Dakota, South Dakota, New Mexico, Louisiana, Vermont, Wisconsin, but also individual counties/boroughs in other states (e.g. part of Long Island, New York City) are affected by these issues. See Uhl et al. (2021) for details.

- As described before, the age estimates from model 1 allow to create topologically intact, and connected depictions of historical road networks, but may result in estimating the construction periods as too early. Models M2 and M3 aim to mitigate this issue but may provide more disconnected depictions of the historical road networks.

- Deconstruction or re-construction of road networks using a different layout (e.g., replacing a grid-like road network with curved roads, or replacements of 4-way intersections by roundabouts) are not reflected in CHRONEX-US.

- CHRONEX-US aims to estimate the age of local/residential roads only. The underlying model assumptions are not valid for interstates or highways. While the method has been applied to all road categories in the NTD dataset, it is advised to exclude non-local roads from analyses. This can be done using the MTFCC attribute.

- In peri-urban areas of low building density, such as agriculturally used or low-density residential areas, model assumptions may not be valid, and estimated road construction epochs may have higher levels of uncertainty.

- CHRONEX-US is of limited suitability for street or street-segment level analysis, as the estimates for individual road segments may be incorrect. Instead, the data should be used for network analyses to assess the general properties of the stock of roads (i.e., the total length of roads existing in a given year) or for large temporal increments (e.g., roads built between 1950 and 1980), as demonstrated in Burghardt et al (2022, 2024a,b).

- CHRONEX-US currently covers the areas delineated by the GBUA dataset and does not include rural areas. In future work, the data coverage will be expanded to include also rural areas in the conterminous United States.

- While HISDAC-US has been validated extensively (Leyk and Uhl 2018, Uhl et al. 2021), formal validation of the GBUA dataset and CHRONEX-US is challenging due to the lack in reference data. However, trends derived from CHRONEX-US (Model 1) have been compared with analytical results derived from other historical models of the US road network (Barrington-Leigh and Millard-Ball 2015, Boeing 2021), and results largely agree (Burghardt et al. 2022).

## Acknowledgments

Funding for this work was provided through the Human Networks and Data Science – Infrastructure program of the US National Science Foundation (Award Numbers 2121976 and 2419335, respectively) to the University of Colorado Boulder. This research benefited from support provided to the University of Colorado Population Center (CUPC, Project 2P2CHD066613-06) from the Eunice Kennedy Shriver Institute of Child Health, Human and Human Development. The content is solely the authors' responsibility and does not necessarily represent the official views of the National Institutes of Health (NIH) or CUPC. We gratefully acknowledge access to the Zillow Transaction and Assessment Dataset (ZTRAX) through a data use agreement between the University of Colorado Boulder and Zillow Group, Inc. The results and opinions are those of the authors and do not reflect the position of Zillow Group. Support by Zillow Group, Inc., is gratefully acknowledged. Moreover, we gratefully acknowledge support by Safe Software, Inc., for providing a Feature Manipulation Engine (FME) license.